\documentclass[aip,reprint,twocolumn,amsmath,amssymb]{revtex4-1}
\usepackage{amsmath}
\usepackage{amsfonts}
\usepackage{color}
\usepackage{braket}
\usepackage{graphicx}% Include figure files
\usepackage{dcolumn}% Align table columns on decimal point
\usepackage{bm}% bold math
\usepackage{prettyref}
\newrefformat{eq}{(\ref{#1})}
\newrefformat{fig}{\ref{#1}}
\def\pref#1{\prettyref{#1}}

\begin{document}

%\preprint{AIP/123-QED}

\title{Proposal of highly efficient photoemitter with strong photon-harvesting capability and exciton superradiance}% Force line
% breaks with \\
%\thanks{Footnote to title of article.}

\author{Takuya Matsuda}
\email{matsuda@pe.osakafu-u.ac.jp}
\affiliation{ 
Department of Physics and Electronics, 
Osaka Prefecture University,
Sakai, Osaka 599-8531, Japan}
% \altaffiliation[Also at ]{Department of Physics and Electronics, Osaka Prefecture University, Sakai, Osaka 599-8631,Japan}%Lines break automatically or can be forced with \\
\author{Hajime Ishihara}
\affiliation{ 
Department of Physics and Electronics, 
Osaka Prefecture University,
Sakai, Osaka 599-8531, Japan}
\affiliation{ 
Department of Materials Engineering Science,  
Osaka University,
Toyonaka, Osaka 560-8531, Japan%\\This line break forced with \textbackslash\textbackslash
}

\vspace{5mm}
%\date{today}% It is always \today, today,
\begin{abstract}
We propose a system of highly efficient photoemitters comprising metal-molecule multilayered structures. In the proposed
structure, the absorption in the molecular layer is greatly enhanced through quantum interference between the split modes 
arising from the coupling of the layered excitons and the plasmons sustained by the metal layer. 
Furthermore, the large interaction volume between surface plasmons and excitons causes exciton superradiance, 
which results in the extremely efficient photoemission. 
This finding indicates the possibility of designing highly efficient photoemitters based on simple layered structures.
\end{abstract}

\maketitle

%%%%%%%%%%%%%%%%%%%%%
\def\dd{{\rm d}}
\def\ee{{\rm e}}
\def\ii{{\rm i}}
\def\ddt{\frac{\partial}{\partial t}}

\def\const{\text{const.}}
\def\cc{\text{c.c.}}
\def\Hc{\text{H.c.}}
\def\Re{\text{Re}}
\def\Im{\text{Im}}
\def\Tr{\text{Tr}}
\def\vnabla{\bm{\nabla}}
\def\rot{\vnabla\times}
\def\eps{\varepsilon}
\def\epsz{\varepsilon_0}
\def\epsb{\varepsilon_{\text{bg}}}
\def\chiex{\chi_{\text{ex}}}
\def\thickex{L_{\text{ex}}}
\def\thick{L}

\def\w{\omega}
\def\wex{\varOmega_\text{ex}}
\def\win{\varOmega_\text{in}}
\def\ww{\varOmega}
\def\mz{m_0}
\def\mex{m_{\text{ex}}}
\def\dampex{{\it \Gamma}_{\text{ex}}}

\def\eV{\text{eV}}
\def\meV{\text{meV}}
\def\ps{\text{ps}}
\def\nm{\text{nm}}

%%%%%%%%%
% vector
%%%%%%%%%
\def\vr{\bm{r}}
\def\vk{\bm{k}}
\def\vrp{\bm{r}_\parallel}
\def\vkp{\bm{k}_\parallel}
\def\vPex{\bm{P}_{\text{ex}}}
\def\vP{\bm{P}}
\def\vE{\bm{E}}
\def\vEz{\bm{E}_0}
%%%%%%%%%%%%%%%%%%%%%%%
% coefficient
%%%%%%%%%%%%%%%%%%%%%%%
\def\vdimI{\bm{\mathcal{I}}}
\def\dimP{\mathcal{P}}
\def\vdimP{\bm{\mathcal{P}}}
\def\vdimPT{\bm{\mathcal{P}}^{\text{T}}}
\def\vdimPL{\bm{\mathcal{P}}^{\text{L}}}
\def\vdimE{\bm{\mathcal{E}}}
\def\vdimF{\bm{\mathcal{F}}}
%%%%%%%%%%%%%%%%%%%%%
% matrix
%%%%%%%%%%%%%%%%%%%%%
\def\munit{\bm{\mathsf{I}}}
\def\mzero{\bm{\mathsf{0}}}
\def\mG{\bm{\mathsf{G}}}
%%%%%%%%%%%%%%%%%%%%%%%%%%%
% operator
%%%%%%%%%%%%%%%%%%%%%%%%%%%
\def\oH{\hat{H}}
\def\oHex{\hat{H}_{\text{ex}}}
\def\oHI{\hat{H}_{\text{I}}}
\def\oex{\hat{b}}
\def\oexd{\hat{b}^{\dagger}}
\def\ovPex{\hat{\bm{P}}_{\text{ex}}}
\def\oPex{\hat{P}_{\text{ex}}}
\def\ovE{\hat{\bm{E}}}
\def\ovEz{\hat{\bm{E}}_{0}}
%%%%%%%%%%%%%%%%%%%%%%%%%%%%%%%

%%%%%Intro%%%%%%%%%%

For the realization of high-efficiency solar cells, photo-emitting devices, and so forth, smart designs of the light-matter coupling 
are necessary~\cite{tang87,baluschev06,riesen16,peng06,reinke06,nowy08,hayashi10}. 
In particular, the scheme of high-efficiency photon harvesting is crucial 
for the development of energy-saving technologies. 
For example, a high concentration of harvested light energy in the photoactive parts of devices is important for realizing efficient photonic functions. One approach to achieving the high concentration
of harvested light energy is to utilize nanometallic structures sustaining surface plasmon (SP) resonance~\cite{wiederrecht04, kuhn06,ueno08,ishi11,yano13,osaka14}.  
It has been proposed that the quantum interference between the split coupled modes comprising SPs and molecular excitons
leads to strong energy concentration in molecules~\cite{ishi11,yano13,osaka14}, which has been discussed as a type of Fano resonance~(FR) appearing through the interference between the sharp discrete resonance and much broader resonance~\cite{luk10}. However, such effects usually require highly sophisticated designs and the fabrication of nanometallic structures 
to ensure strict control of the system parameters.
Thus, it is desirable to realize systems for efficient photon-harvesting with simpler designs such as
those with layered structures.
In studies on the control of the light-matter interaction, it was reported that metal-molecular composite layered structures easily induce large Rabi splitting~(RS) due to the strong interaction between excitons and SPs~\cite{bellessa04,bonnand06,hakala09,cade09}. Also, Hayashi {\it et al}. proposed highly sensitive sensors based on the interference between the SP mode and waveguide mode in multilayered structures~\cite{hayashi16}. However, such structures have received little interest for application to photon-harvesting through the mode interference effect~(Fano resonance).

On the other hand, to realize highly efficient photoemitters, not only an efficient photon-harvesting scheme but also an efficient photoemission scheme must be realized. It is known that highly efficient photoemission occurs through exciton superradiance. 
As the interaction volume (or coherence volume) between excitons and the radiation field increases, the radiative decay rate increases (so-called exciton superradiance)~\cite{feldmann87, nakamura89, itoh90, knoester92,bjork95,agranovich97}. For example, for very high quality CuCl thin films, the radiative decay time was observed to reach 100~fs order~\cite{ichimiya09}, where excitons with a very high coherence volume over the entire sample are strongly coupled with the radiation field. 

%%%%%Intro//%%%%%%%%%%

If we can design simple layered structures in which energy concentration and exciton superradiance are simultaneously realized, it will greatly contribute to the realization of photoemitters with high efficiency. In this paper, we demonstrate a condition under which strong energy concentration occurs in metal-molecule composite multilayered structures, where SP-induced exciton superradiance occurs simultaneously. 

% CCCCCCCCCCCCCCCCCCCCCCCCCCCCCCCCCCCCCCCCCCCCCCCCCCCCCCCCCCCC
\begin{figure}[htbp] % CCCCC
%\vspace{-1em}
\includegraphics[width=60mm]{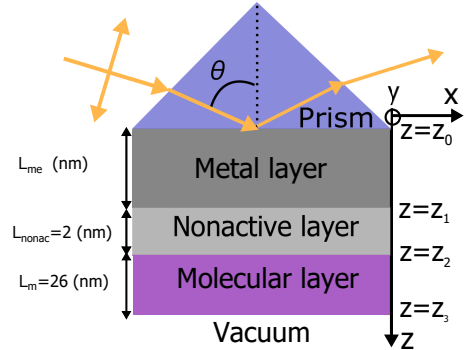}
\caption{Sketch of the assumed composite layered structure and $p$-polarized incident light is assumed to excite the SP modes.}
\label{fig1}
\end{figure} 
% CCCCCCCCCCCCCCCCCCCCCCCCCCCCCCCCCCCCCCCCCCCCCCCCCCCCCCCCCCCC

We consider a Kretschmann configuration consisting of a prism, a metal layer, a nonactive layer, a molecular layer, and a vacuum as illustrated in Fig.~\ref{fig1}.
Unless otherwise noted, we use the same parameters as in Ref.~\onlinecite{bonnand06}. We treat a prism as a semi-infinite layer. The nonactive layer is TiO$_{2}$ and its thickness $\thick_{\rm nonac}$ is $2~\nm$. We assume that the molecular layer with thickness $\thick_{\rm m}=26~\nm$ is a PVA-TDBC one. This structure enables SP modes to propagate along the vacuum-side surface of the molecular layer. The dielectric function of the metal layer is modeled by the three critical point pole pairs (CP3) model with the parameters of Ag~\cite{lu10}. 

We consider the following Hamiltonian: $\oH=\oHex+\oHI$, where $\oHex=\hbar\wex\oexd\oex$ represents the resonance energy of a bare exciton in the molecular layer and $\hbar\wex=2.1~\eV$. The bosonic operator $\oex~(\oexd)$ stands for the annihilation (creation) operator of an exciton. $\oHI= -\int\dd\vr~\ovPex(\vr)\cdot\vE(\vr,t)$ represents the interaction between excitons and the electric field. The excitonic polarization is represented as $\ovPex(\vr)=\vdimP(\vr)\oex+\Hc$, where the expansion coefficient $\vdimP(\vr)$ is expressed as $\vdimP(\vr)=\vP~\ee^{\ii\vkp\cdot\vrp}\big\{{\it \Theta}(z-z_{j-1})-{\it \Theta}(z-z_{j})\big\}$, where $\vP$ is the transition dipole moment density, and $\vkp$ and $\vrp$ are a wavevector and position vector parallel to the film surface, respectively. ${\it \Theta}(z)$ is the Heaviside step function. Here, we consider only the excitonic polarization as the coordinate phase mode, and the relative motion of excitons is treated as that in the bulk. We express the electric field satisfying the Maxwell equation in integral form as
\begin{align}\label{eq:Maxwell}
\vE(\vr,\w)
=
\vEz(\vr,\w)+\int\dd\vr^{\prime}~\bar{\mG}(\vr,\vr^{\prime},\w)\cdot\vdimP(\vr^{\prime})\braket{\oex(\w)}.
\end{align}
The dyadic Green's function $\bar{\mG}(\vr,\vr^{\prime},\w)$ satisfies the equation $\big[\rot\rot-\eps(\vr,\w)\w^{2}/c^{2}\big]\bar{\mG}(\vr,\vr^{\prime},\w)=\w^{2}/(\epsz c^{2})\delta(\vr-\vr^{\prime})~\bar{\munit}$~\cite{chew95}, where $\eps(\vr,\w)$ reflects the sample geometry determined by the background dielectric constants of the different layers. In this expression, $c$ is the speed of light and $\epsz$ is the vacuum permittivity. Solving the equation of motion for excitons and the Maxwell equation simultaneously, we obtain the self-consistent equation set as
\begin{subequations}
\begin{align}
\label{eq:bex}
&\big[\hbar(\wex-\w-\ii\dampex/2)+\mathcal{A}(\w)\big]
\braket{\oex(\w)}
\nonumber\\
&=
\int\dd\vr~\vdimP^{\ast}(\vr)\cdot\vEz(\vr,\w),
\\
\label{eq:rad}
\mathcal{A}(\w)
&\equiv
-\int\dd\vr\int\dd\vr^{\prime}~
\vdimP^{\ast}(\vr)\cdot\bar{\mG}(\vr,\vr^{\prime},\w)\cdot\vdimP(\vr^{\prime}),
\end{align}
\end{subequations}
where $\vEz(\vr,\w)$ is the incident electric field. In this expression, we phenomenologically introduce nonradiative damping $\dampex$ whose value is $49~\meV$. Equation \pref{eq:rad} describes the energy correction of excitons including the effect of SPs through $\bar{\mG}(\vr,\vr^{\prime},\w)$. We can now determine the self-consistent field by substituting $\braket{\oex(\w)}$ from Eq.~\pref{eq:bex} into Eq.~\pref{eq:Maxwell}.

First, we investigate the absorptivity spectrum in the molecular layer while varying the thicknesses of the metal layer. We evaluate the absorptivity spectrum in the $j$th layer, $A_{j}(\w)~(j \in\mathbb{N})$, using the following expressions~\cite{kim12}:
$A_{j}(\w)
\equiv
\frac{1}{S_{0}(\w)}
\int_{z_{j-1}}^{z_{j}}\dd z~
Q_{j}(z,\w)$, where $Q_{j}(z,\w)$ is the optical power dissipation of the $j$th layer in the $z$ direction. $S_0(\w)=(1/2)\Re\{E_{x,0}H_{y,0}^\ast\}$ is the magnitude of the input time-averaged Poynting vector, where $E_{x,0}$ is the $x$ component of the electric field in the input region and $H_{y,0}$ is the $y$ component of the magnetic field in the input region. If the absorptivity at a particular layer is 1, it means that an incident light power is totally concentrated into this layer without reflectance and transmittance.

% CCCCCCCCCCCCCCCCCCCCCCCCCCCCCCCCCCCCCCCCCCCCCCCCCCCCCCCCCCCC
\begin{figure}[htbp] % CCCCC
%\vspace{-1em}
\includegraphics[width=\linewidth]{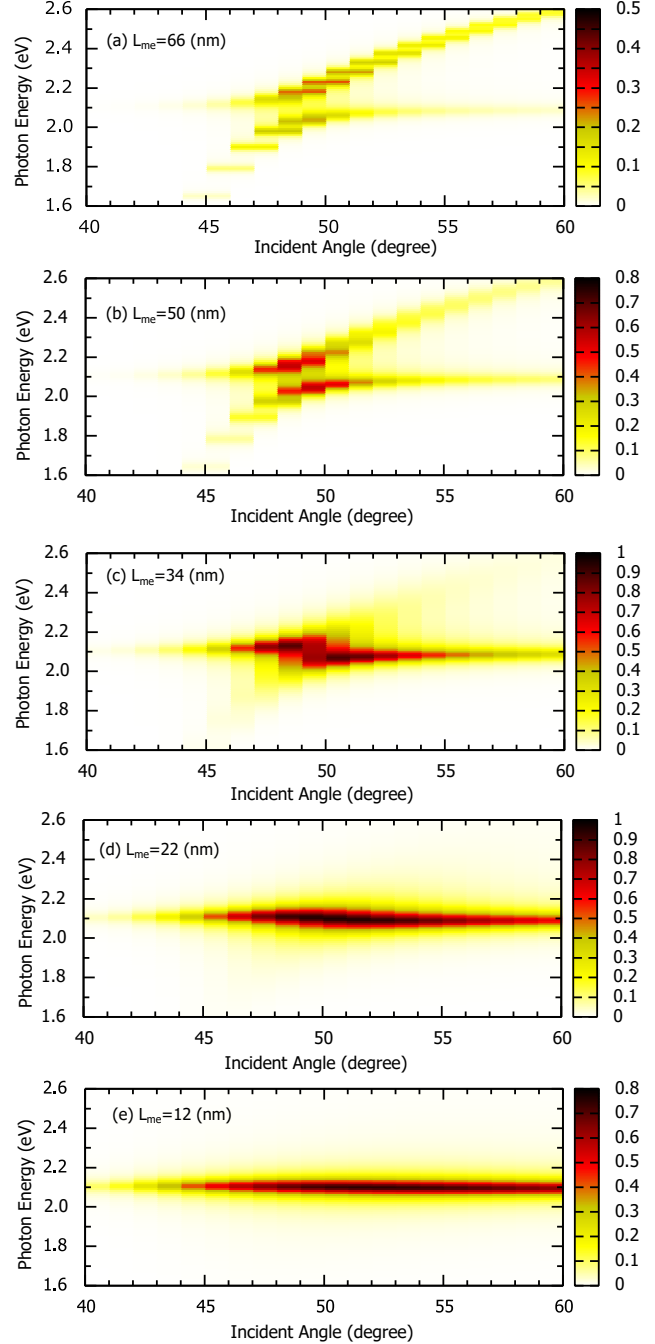}
\caption{Absorptivity in the molecular layer plotted as a function of the incident angle and incident photon energy for different thicknesses of the metal layer: (a) $\thick_{\rm me}=66~\nm$, (b) $\thick_{\rm me}=50~\nm$, (c) $\thick_{\rm me}=34~\nm$, (d) $\thick_{\rm me}=22~\nm$, and (e) $\thick_{\rm me}=12~\nm$.}
\label{fig2}
\end{figure} % CCCCC
% CCCCCCCCCCCCCCCCCCCCCCCCCCCCCCCCCCCCCCCCCCCCCCCCCCCCCCCCCCCC

Here, we assume that the $p$-polarized incident light is used to excite the SP modes. 
In Fig.~\ref{fig2}, we plot the absorptivity in the molecular layer as a function of the incident angle and incident photon energy. In Figs.~\ref{fig2}(a)-(e), we confirm the crossover behaviour from RS to FR with decreasing thickness of the metal layer. As shown in Figs.~\ref{fig2}(a,b), in the RS regime the absorption by the two modes in the molecular layer is enhanced at particular incident angles. On the other hand, in the FR regime, the absorption of only one of the modes in the molecular layer is greatly enhanced over a wide range of the incident angle as shown in Figs.~\ref{fig2}(d,e). Within the thickness regime in our system, the thicker metal film generates a larger plasmonic dipole moment that leads to the stronger coupling with the molecular excitons. On the other hand, with the decrease in the metal film thickness, the coupling strength between the plasmons and excitons rapidly decreases. Also, note that, in FR regime, we do not need to carefully choose the incident angle to enhance the absorption in the molecular layer.

To find suitable conditions to enhance the absorption in the molecular layer, we examine the following two cases: (i) the thickness of the metal layer $\thick_{\rm me}$ is $50~\nm$ and the incident angle $\theta$ is $49^{\circ}$ in the RS regime, and (ii) $\thick_{\rm me}=22~\nm$ and the incident angle $\theta=50^{\circ}$ in the FR regime.

% CCCCCCCCCCCCCCCCCCCCCCCCCCCCCCCCCCCCCCCCCCCCCCCCCCCCCCCCCCCC
\begin{figure}[htbp] % CCCCC
%\vspace{-1em}
\includegraphics[width=\linewidth]{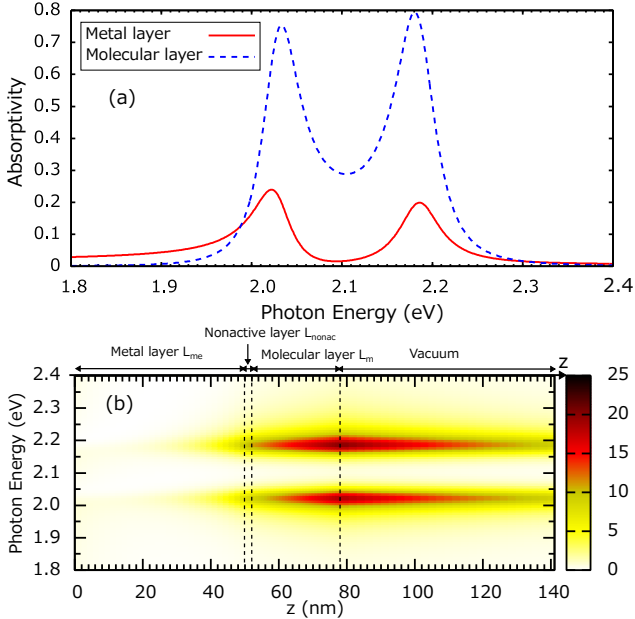}
\caption{(a) Absorptivity spectra for structures with thickness of the metal layer $\thick_{\rm me}=50~\nm$ and incident angle $\theta=49^{\circ}$ plotted as a function of the incident photon energy. The red solid line represents the absorption in the metal layer and the blue dashed line represents that in the molecular layer. (b) Electric field intensity as a function of position $z$ and incident photon energy.}
\vspace{-1em}
\label{fig3}
\end{figure} % CCCCC
% CCCCCCCCCCCCCCCCCCCCCCCCCCCCCCCCCCCCCCCCCCCCCCCCCCCCCCCCCCCC

In case (i) (RS regime), as shown in Fig.~\ref{fig3}(a), the enhancement of absorption in the metal layer cannot be avoided at the peak energy of the absorption in the molecular layer because the Rabi-split modes contain both plasmon and exciton components owing to their strong coupling. Figure~\ref{fig3}(b) shows the electric field intensity as a function of position $z$ and incident photon energy. As shown in Fig.~\ref{fig3}(b), the electric field is enhanced at the peak positions of absorption in Fig.~\ref{fig3}(a). Thus, the RS regime is not suitable for efficient photon-harvesting because the incident photon energy cannot be concentrated into only the molecular layer.

% CCCCCCCCCCCCCCCCCCCCCCCCCCCCCCCCCCCCCCCCCCCCCCCCCCCCCCCCCCCC
\begin{figure}[htbp] % CCCCC
%\vspace{2em}
\includegraphics[width=\linewidth]{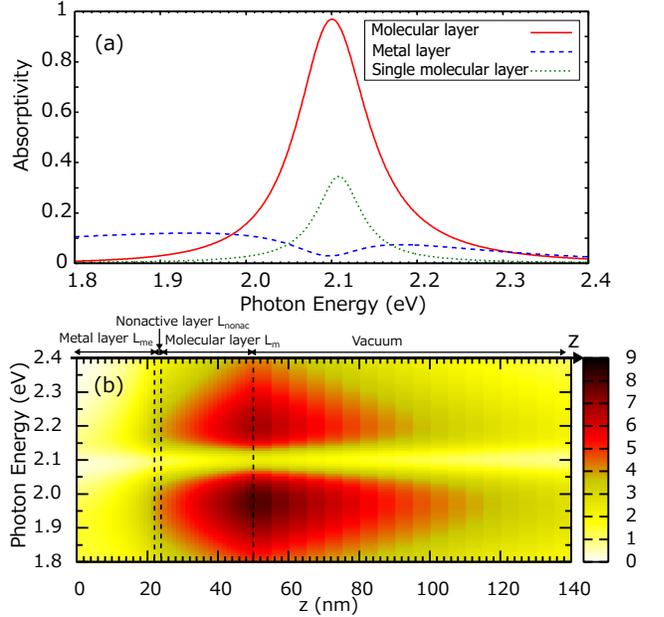}
\caption{(a) Absorptivity spectra for structures with thickness of the metal layer $\thick_{\rm me}=22~\nm$ and incident angle $\theta=50^{\circ}$ plotted as a function of the incident photon energy. The red solid line represents the absorption in the metal layer and the blue dashed line represents that in the molecular layer. The green dotted line shows the absorption in a single molecular layer (for the optimum condition; incident angle of $s$-polarized light of $\theta=80^{\circ}$) for comparison with the composite layered structure with $\thick_{\rm me}=22~\nm$ and $\theta=50^{\circ}$. (b) Electric field intensity plotted similarly to that in Fig.~\ref{fig3}(b).}
\label{fig4}
\end{figure} % CCCCC
% CCCCCCCCCCCCCCCCCCCCCCCCCCCCCCCCCCCCCCCCCCCCCCCCCCCCCCCCCCCC

In case (ii) (FR regime), we find the condition of strong suppression of the absorption in the metal layer ($2\%$) as shown by the red solid line in Fig.~\ref{fig4}(a), and, at the same frequency, marked enhancement of the absorption in the molecular layer ($96\%$) shown by the blue dashed line in Fig.~\ref{fig4}(a). This value is considerably superior to that for the system in Ref.~\onlinecite{ishi11}. On the other hand, in the single molecular layer [see the green dotted line in Fig.~\ref{fig4}(a)], the absorptivity does not reach $40\%$ even under the optimum condition, the incident angle of $s$-polarized light of $\theta=80^{\circ}$ for the same thickness. Figure~\ref{fig4}(b) shows the electric field profile plotted similarly to that in Fig.~\ref{fig3}(b). This profile indicates that excitons are not very strongly coupled with SPs, and the energy concentration occurs in the molecular layer~\cite{ishi11} because, in the FR regime, the splitting between exciton-plasmon coupled modes is not larger than the spectral width of the respective modes, and at the excitonic resonance energy, these two peaks are overlapped, where the excitonic components are constructively superposed, while the plasmonic components destructively superposed. This quantum interference clearly appears in Fig 4(a). Namely, the plasmonic absorption is strongly suppressed and the excitonic resonance is remarkably enhanced. 
(Note that the electric field intensity at around 2.1 eV is suppressed at the molecular layer in spite of the greatly enhanced absorptivity because of the interference between the incident field and the strong radiated field from the induced polarization in layered structures.) 
In this way, strong energy concentration into the molecular layer alone is possible in the FR regime, which is advantageous for realizing efficient photon-harvesting with simple layered structures. 

%%%%%%%%%%%%%%%%%%%%%%%%%%%%%%%%%%%%%%%%%%%%%%%

Next, we examine the possibility of exciton superradiance in our proposed structure. To evaluate the radiative decay time of excitons, we find the roots $\{\tilde{\w}\}$ of ${\rm det}[\hbar(\wex-\tilde{\w})+\mathcal{A}(\tilde{\w})]=0$ in the LHS of Eq.~\pref{eq:bex}. Here, we set the nonradiative damping included in the dielectric function of Ag to zero for the technical reason to extract the pure radiative width of excitons. In fact, the superradiance is hardly affected by the the nonradiative damping of Ag
because the plasmonic excitation is strongly suppressed in the considered condition.  
The roots $\{\tilde{\w}\}$ provide the complex eigenfrequencies $\{\tilde{\w}\}$ of the coupled mode, whose real parts $\Re\{\tilde{\w}\}$ give the eigenfrequencies including the radiative shift and whose imaginary parts $\Im\{\tilde{\w}\}$ correspond to the radiative decay rate. Thus, we can describe the radiative decay time as $\tau_{\rm R}=1/(-2\Im\{\tilde{\w}\})$.

By evaluating $\tau_{\rm R}$, we find that the radiative decay time of excitons in our proposed structure ($\tau_{\rm R}$=0.012~\ps) is much shorter than that in the single layered structure with the optimum condition, the incident angle of $s$-polarized light of $\theta=80^{\circ}$ ($\tau_{\rm R}$=0.046~\ps), because of the large interaction volume between coherently extended wavefunction of excitons and SPs. These results indicate that our proposed structure has the major advantage of highly efficient photoemitters owing to the compatibility of the strong absorption and large radiative width. 

\begin{figure}[htbp] % CCCCC
\vspace{2em}
\includegraphics[width=\linewidth]{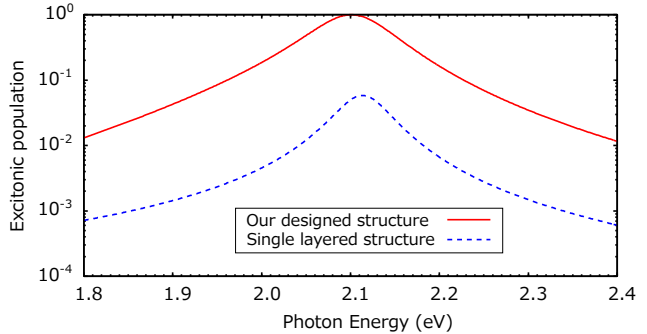}
\caption{(Color online) Excitonic population spectrum versus incident photon energy. Both spectra are normalized by the peak of the red solid line.}
\label{fig5}
\end{figure}

Finally, we calculate excitonic population spectrum to discuss the amount of luminescence in the molecular layer within the linear response regime. 
The excitonic population spectrum can be expressed as~\cite{ishikawa13}
\begin{align}
N_\text{ex}(\w)
&=
|\braket{\oex(\w)}|^2
\nonumber\\
&=
\Big|
\frac{\int\dd\vr~\vdimP^{\ast}(\vr)\cdot\vEz(\vr,\w)}{\hbar(\wex-\w-\ii\dampex/2)+\mathcal{A}(\w)}
\Big|^2.
\end{align}
In Fig.~\ref{fig5}, we plot the excitonic population spectrum versus the incident photon energy. By calculating the integrated intensity of each excitonic population spectrum, we find that the intensity in the present structure is about 24 times larger than that of the single molecular layer under the optimum condition, the incident angle of $s$-polarized light of $\theta=80^{\circ}$. From these results, we can expect marked strengthening of the luminescence utilizing our proposed structure.

% CCCCCCCCCCCCCCCCCCCCCCCCCCCCCCCCCCCCCCCCCCCCCCCCCCCCCCCCCCCC
 % CCCCC
% CCCCCCCCCCCCCCCCCCCCCCCCCCCCCCCCCCCCCCCCCCCCCCCCCCCCCCCCCCCC

%%%%%%%%%%%%%%%%%%%%%%%%%%%%%%%%%%%%%%%%%%%%%%%%%%%%%%%%%%%%%%%%%%%%%%%%%%%%%%%
%
To conclude, we find that high-efficiency energy concentration in the photoactive part of a device can be realized with a simple metal-molecule multilayered structure, where the absorption at the metal layer is greatly suppressed by the Fano resonance effect even for this simple structure. We propose a high-efficiency photoemitter using this mechanism, where strong light-harvesting by the above mechanism and strong photoemission by exciton superradiance occur simultaneously. To demonstrate the potential of our proposed structure, we numerically 
calculate the absorption, radiative decay rate, and population of a layered exciton under particular conditions. As a result, we find a condition under which every aspect commonly shows excellent performance. We hope that our findings will enable the design of high-efficiency photoemitters based on simple layered structures.

%%%%%%%%%%%%%%%%%%%%%%%%%%%%%%%%%%%%%%%%%%%%%%%%%%%%%%%%%%%%%%%%%%%%%%%%%%%%%%%

We thank T. Yano and N. Yokoshi for helpful discussions. This work was partially supported by Grant-in-Aid for JSPS Fellows No.16J11326 from MEXT, Japan, and by JSPS KAKENHI Grant No.JP16H06504 in Scientific Research 
on Innovative Areas ``Nano-Material Optical Manipulation''.
%%%%%%%%%%%%%%%%%%%%%%%%%%%%%%%%%%%%%%%%%%%%%%%%%%%%%%%%%%%%%%%%%%%%%%%%%%%%%%%

%%%%%%%%%%%%%%%%%%%%%%%%%%%%%%%%%%%%%%%%%%%%%%%%%%%%%%%%%%%%%%%%%%%%%%%%%%%%%%%

\end{document}